\documentstyle[12pt]{article}
\catcode`\@=11
\def\makepreprititle{\par
  \begingroup
  \def\thefootnote{\fnsymbol{footnote}}
  \def\
@makefnmark{\hbox
  to 0pt{$^{\@thefnmark}$\hss}}
  \if@twocolumn
  \twocolumn[\@makepreprititle]
  \else \newpage
  \global\@topnum\z@
  \@makepreprititle \fi\thispagestyle{empty}\@thanks
  \endgroup
  \setcounter{footnote}{0}
  \let\makepreprititle\relax
  \let\@makepreprititle\relax
  \gdef\@thanks{}\gdef\@author{}\gdef\@title{}
  \gdef\@preprintnumber{}\gdef\@preprintdate{}\gdef\subtitle{}
  \let\thanks\relax}
\def\preprintnumber#1{\gdef\@preprintnumber{#1}}
\def\preprintdate#1{\gdef\@preprintdate{#1}}
\def\subtitle#1{\gdef\@subtitle{#1}}
\def\@makepreprititle{\newpage
{\def\baselinestretch{1}
  \begin{flushright} \@preprintnumber \par
  \@preprintdate \end{flushright} } \par
  \begin{center}
\vskip 1.5em
  {\LARGE \@title \par} \vskip 2.5em
  {\Large \lineskip .5em
  \begin{tabular}[t]{c}\@author
  \end{tabular}\par}
  \vskip 1em {\large \@date} \end{center}
  \par
  \vfil}
\date{\sl Department of Physics, Tohoku University\\Sendai, 980 Japan}
\preprintdate{~}
\preprintnumber{~}
\subtitle{}
\def\abstract{\if@twocolumn
\section*{Abstract}
\else \normalsize
\begin{center}
{\bf Abstract\vspace{-.5em}\vspace{0pt}}
\end{center}
\quotation
\addtocounter{page}{-1}
\fi}
\def\endabstract{\if@twocolumn\else\endquotation\fi}
\def\spacing#1{\def\baselinestretch{#1}
\typeout{baselinestretch is modified to \baselinestretch}}
\catcode`\@=12

\spacing{1.4}
\title{The Weinberg Angle\\Without Grand Unification}
\author{T.~Moroi, Hitoshi Murayama and T.~Yanagida}
\date{\sl Department of Physics, Tohoku University\\
Sendai, 980 Japan}
\preprintnumber{TU-438}
\preprintdate{June, 1993}
\begin{document}
\makepreprititle
\begin{abstract}
We assume that strong and electroweak interactions become strong at very
high energies. With this assumption, we compute the low-energy gauge
coupling constants $\alpha_i (m_Z)$ as a function of the cutoff scale,
taking the supersymmetric standard model with $3+2n$ families of quark and
lepton multiplets. We find that only the five family case $(n=1)$ is
consistent with the experimental values of the gauge coupling constants.
This suggests the presence of a pair of families at $\sim 1$~TeV.
\end{abstract}

\newpage

Supersymmetry (SUSY) \cite{SUSY} has attracted many theorists in
particle physics for a long time, since it is not only a mathematically
consistent symmetry, but also it can eliminate all divergences more than
quadratic in renormalizable quantum field theories. At present, the SUSY
is expected to be a solution to the hierarchy problem (the presence of
the light Higgs boson) \cite{naturalness}. However, in the SUSY
extension of the standard electroweak theory with three families of
quarks and leptons, the electroweak gauge interactions are not
asymptotically free, and hence it may not be a consistent theory for all
energies \cite{Wilson}.

The grand unification (GUT) \cite{GUT} of strong and electroweak
interactions is certainly a possible solution to this problem, since all
the gauge interactions become asymptotically free above the GUT scale.
Furthermore, this SUSY-GUT \cite{SUSY-GUT} is strongly supported
phenomenologically by the recent measurement of $\sin^2 \theta_W$ made
at the LEP experiments \cite{LEP}. The SUSY-GUT, however, has a serious
problem to which no convincing solution has been found. Namely, the
extreme fine tuning of parameters is required to produce a large mass
splitting in Higgs multiplets. It is, therefore, very important to
pursue alternative solutions to the problem of the non-asymptotically
free nature of the electroweak interactions.

In the present paper we examine the proposal by Parisi \cite{Parisi}, in
which a physical cutoff $\Lambda$ is introduced such that the
non-asymptotically free theory is valid up to the scale $\Lambda$. If
the cutoff $\Lambda$ is sufficiently large, the gauge coupling constants
at low energy (at the renormalization scale $\mu \sim 100~{\rm GeV}$)
must be near the infrared stable fixed point $\alpha_i = 0$. Moreover,
if the coupling constants at $\Lambda$ are large enough, the low-energy
coupling constants are approximately independent of the values of
couplings at $\Lambda$ and are determined solely in terms of the cutoff
$\Lambda$ and the number of matter multiplets \cite{Parisi}. Notably 30
years ago, Landau \cite{Landau} suggested that the presence of a large
number of fermions could give an explanation of the small value of the
fine-structure constant $\alpha_{em} = 1/137$. A similar possibility has
also been stressed based on a composite model of the gauge bosons
\cite{composite}.

In this paper we asuume that SUSY standard model with $3+2n$ families of
quarks and leptons. We take the cutoff scale $\Lambda$ as a free
parameter which will be determined as a scale where all gauge coupling
constants blow up. This is a crucial point to make our model viable
phenomenologically, while in all previous analyses
\cite{MPP,Parisi,Landau,ACMT} the cutoff $\Lambda$ is taken at the
Planck scale, and hence they face a difficulty to obtain the correct
value of the Weinberg angle \cite{MY}.

To determine $\Lambda$ we compute the low-energy coupling constants
$\alpha_1 (m_Z)$, $\alpha_2 (m_Z)$ and $\alpha_3 (m_Z)$ as a function of
$\Lambda$ using the two-loop renormalization group (RG) equations
\cite{RG}. We assume that the extra $2n$ families form $n$ pairs of a
family and a mirror family (family pairs) so that they can have $SU(2)
\times U(1)$ invariant masses $m_F$. In this case, neutrinos in the
extra families become massive Dirac fermions of the mass $m_F$ escaping
the constraints from the LEP experiments. The results are shown in
Fig.~1, for the number of family pairs $n=1,\,2,\,3$. In this Figure, we
have taken $m_F = m_{SUSY} = 1$~TeV where $m_{SUSY}$ is the
SUSY-breaking scale. Note that $3+2n=5$ is the minimum number of
families which makes all the gauge coupling constants non-asymptotically
free. To reproduce the correct values for the three low-energy gauge
coupling constants, we find that the minimal number of the family pairs
$n=1$ is the unique choice, and the cutoff turns out to be $\Lambda
\simeq 2 \times 10^{16}$~GeV
\cite{3-loop}. Here we have taken $\alpha_1 (\Lambda) = \alpha_2
(\Lambda)= \alpha_3 (\Lambda)= 10$ as the initial value of the RG equations.
We will restrict our discussions to the case $n=1$ hereafter.

To see that the low-energy gauge coupling constants are in fact
independent of their values at the cutoff scale $\Lambda$, we vary the
initial $\alpha_i$ between 10 and 100. Fig.~2 shows the running of the
Weinberg angle $\sin^2 \theta_W$ just below the cutoff $\Lambda$. The
initial values of $\sin^2 \theta_W$ are varied between 0.05 to 1.0, and
they rapidly approach the value $\sin^2 \theta_W \simeq 3/8$ around
$10^{16}$~GeV. The further running down to the electroweak scale
naturally reproduces the experimental value $\sin^2 \theta_W (m_Z) =
0.2326 \pm 0.0008$, just as in the conventional SUSY-GUT. We have
checked that even when $\alpha_1, \alpha_2$, and $\alpha_3$ are taken
from 1 to 100 at the cutoff scale, their low-energy values scatter only
within 2~\% for $\alpha_1 (m_Z)$, 5~\% for $\alpha_2 (m_Z)$ and 10~\%
for $\alpha_3 (m_Z)$. Thus, this scenario predicts all the low-energy
gauge coupling constants almost irrespective of the initial values of
the large gauge coupling constants at the cutoff scale.

The predicted gauge coupling constants are sensitive on the mass of the
family pair as well as the SUSY-breaking scale, while insensitive on
the initial gauge coupling constants. Varying $m_F = m_{SUSY}$ between
100~GeV and 10~TeV, we first determine the cutoff scale $\Lambda$ so
that the correct weak-scale fine-structure constant $\alpha_{em}^{-1}
(m_Z)= 127.9 \pm 0.2$ is obtained. With the determined $\Lambda$ we
compute the Weinberg angle $\sin^2 \theta_W (m_Z)$ and the QCD running
coupling $\alpha_3 (m_Z)$ at the weak scale.  The results are shown in
Fig.~3. We see that the present model predicts one pair of a family and
a mirror family at around 1~TeV as well as the SUSY particles. Thus, the
present model will be testable at the future supercollider experiments
such as SSC or LHC \cite{mSUSY}.

A potential problem in the present model is the presence of the family
pair at the TeV scale. Since they can have $SU(2) \times U(1)$ invariant
masses, one needs to explain why they are so light compared with the
cutoff scale $\Lambda \simeq 10^{16}$~GeV. A possible solution to this
problem is given if they are Nambu--Goldstone supermultiplets arising
from a spontaneous breakdown of some global symmetry $G$. An example is
$E_6 \rightarrow SO(10) \times U(1)$, where the Nambu--Goldstone
multiplets are {\bf 16} and ${\bf \overline{16}}$ of $SO(10)$ for the
doubling realization case \cite{foot}. In this case the Nambu--Goldstone
supermultiplets naturally have the masses of the order of the
SUSY-breaking scale $m_{SUSY}$ \cite{BLPY}. Thus, this example naturally
explains the presence of a pair of a family and a mirror family at the
SUSY-breaking scale.

The other problem is to clarify the physics at the cutoff scale. Since
the cutoff $\Lambda$ is much lower than the Planck scale, it is
plausible that the underlying physics is independent of the gravity. An
intriguing possibility is that all the gauge fields of $SU(3) \times
SU(2) \times U(1)$ are composite fields generated by the dynamics at the
cutoff scale \cite{ES}. However, the dynamics to generate the composite
gauge fields has not been yet clear to us.

\newpage
\section*{Figure Caption}

\renewcommand{\labelenumi}{Fig.~\arabic{enumi}}
\begin{enumerate}
\item The cutoff dependence of the low-energy gauge coupling constants,
$\alpha_1 (m_Z)$ (Fig.~1a), $\alpha_2 (m_Z)$ (Fig.~1b), and $\alpha_3
(m_Z)$ (Fig.~1c), for various number of the family pairs $n =
1,\,2,\,3$. The masses of the family pairs and SUSY-breaking scale are
taken at 1~TeV. We adopt the gauge coupling constants given by
P.~Langacker, Pennsilvania University preprint, UPR-0492T, (1992).
\item The running of the Weinberg angle $\sin^2 \theta_W$ below the
cutoff $\Lambda = 2 \times 10^{16}$~GeV. The gauge coupling
constants at $\Lambda$ are taken randomly between 10 and 100.
\item The dependence of $\sin^2 \theta_W (m_Z)$ (Fig.~3a) and $\alpha_3
(m_Z)$ (Fig.~3b) on the mass of the family pair $m_F$ and SUSY-breaking
scale $m_{SUSY}$ which are set equal. The cutoff $\Lambda$ is chosen to
reproduce the correct weak-scale fine-structure constant $\alpha_{em}
(m_Z) = 127.9$.
\end{enumerate}


\newpage
\begin{thebibliography}{99}
\bibitem{SUSY} D.~Volkov and V.P.~Akulov, {\sl JETP Lett.}\/ {\bf 16}
(1972) 438;\\
J.~Wess and B.~Zumino, {\sl Nucl. Phys.}\/ {\bf B70} (1974) 39; {\sl
Phys. Lett.}\/ {\bf 49B} (1974) 52.
\bibitem{naturalness}
 M. Veltman, {\sl Acta Phys. Pol.}\/ {\bf B12}, 437 (1981);\\
 L.~Maiani, {\it Gif-sur-Yvette Summer School on Particle Physics},\/ 11th,
 Gif-sur-Yvette, France, 1979 (Inst. Nat. Phys. Nucl. Phys. Particules, Paris,
 1979);\\
 S. Dimopoulos and S. Raby {\sl Nucl. Phys.}\/ {\bf B192}, 353 (1981);\\
 E. Witten, {\sl Nucl. Phys.}\/ {\bf B188}, 513 (1981);\\
 M. Dine, W. Fischler and M. Srednicki, {\sl Nucl. Phys.}\/ {\bf B189}, 575
 (1981).
\bibitem{Wilson} K.~Wilson, {\sl Phys. Rev.}\/ {\bf D3} (1971) 1818.
\bibitem{GUT} J.C.~Pati and A.~Salam, {\sl Phys. Rev.}\/ {\bf D10}
(1974) 275;\\
H.~Georgi and S.~Glashow, {\sl Phys. Rev. Lett.}\/ {\bf 32} (1974) 438.
\bibitem{SUSY-GUT} E.~Witten, in Ref.~\cite{naturalness};\\
S.~Dimopoulos, S.~Raby, and F.~Wilczek, {\sl Phys. Rev.}\/
{\bf D24}, 1681 (1981); \\
S.~Dimopoulos and H.~Georgi, {\sl Nucl. Phys.}\/ {\bf B193}, 150 (1981);\\
N.~Sakai, {\sl Zeit. Phys.}\/ {\bf C11}, 153 (1981).
\bibitem{LEP} P.~Langacker and M.-X.~Luo, {\sl Phys. Rev.}\/ {\bf D44}, 817
(1991);\\
U.~Amaldi, W.~de~Boer and H.~F\"{u}rstenau, {\sl Phys. Lett.}\/ {\bf
260B}, 447 (1991);\\
W.J.~Marciano, Brookhaven preprint, BNL-45999, April 1991.
\bibitem{Parisi} G.~Parisi, {\sl Phys. Rev.}\/ {\bf D11} (1975) 909.
\bibitem{Landau} L.D.~Landau, A.A.~Abrikosov, and I.M.~Khalatnikov, {\sl
Dokl. Akad. Nauk USSR}\/ {\bf 95} (1954) 773, 1177; {\it ibid.}\/ {\bf
96} (1954) 261;\\
L.D.~Landau and I.~Pomeranchuk, {\sl Dokl. Akad. Nauk USSR}\/ {\bf 102}
(1955) 489;\\
L.D.~Landau, ``Niels Bohr and the development of physics,'' ed. W.~Pauli
(Pergamon Press, 1955).
\bibitem{composite} T.~Saito and K.~Shigemoto, {\sl Prog. Theor. Phys.}\/ {\bf
57} (1977) 242;\\
H.~Terazawa, Y.~Chikashige and K.~Akama, {\sl Phys.  Rev.}\/ {\bf D15}
(1977) 480.
\bibitem{MPP} L.~Maiani, G.~Parisi, and R.~Petronzio, {\sl Nucl.
Phys.}\/ {\bf B136} (1978) 115.
\bibitem{ACMT} H.~Terazawa, Y.~Chikashige, K.~Akama and T.~Matsuki, {\sl
Phys. Rev.}\/ {\bf D15} (1977) 1181.
\bibitem{MY} The previous analyses \cite{MPP,Parisi,Landau,composite}
did not include SUSY. To make all the three gauge coupling constants
blow up at a single cutoff scale without SUSY, one has to introduce
other matter multiplets in addition to the extra families in somewhat
artificial manner; see, for example, H.~Murayama and T.~Yanagida, {\sl
Mod. Phys. Lett.}\/ {\bf A7}, (1992) 147.
\bibitem{RG} M.B.~Einhorn and D.R.T.~Jones, {\sl Nucl. Phys.}\/ {\bf
B196}, 475 (1982). We have chosen the normalization of the hypercharge
coupling constant such that $\alpha_1 = \frac{5}{3} \alpha_2 \tan^2
\theta_W$.
\bibitem{3-loop} Since the gauge coupling constants become strong around
the cutoff scale, the higher order effects may be important in
determining the precise value of the cutoff scale $\Lambda$. However, the
low-energy gauge coupling constants are insensitive on the detailed
value of $\Lambda$.
\bibitem{mSUSY} If we allow mass splitting in the family pair as well as
in the SUSY spectrum, it is possible that some of the extra particles in
the family pair are lighter than 1~TeV.
\bibitem{foot} In the doubling realization of $G$, there arise
Nambu-Goldstone chiral multiplets $\Phi^i$ corresponding to each broken
generators $T^i$. Since the scalar field of a chiral multiplet is a
complex field, the number of the massless bosons are doubled. Half of
them are true Nambu-Goldstone bosons and the other half are so-called
quasi Nambu-Goldstone bosons. For details see W.~Buchm\"uller, R.~Peccei
and T.~Yanagida, {\sl Nucl. Phys.}\/ {\bf B227} (1983) 503.
\bibitem{BLPY} Once the SUSY is broken at $m_{SUSY}$, the fermion fields
of the Nambu--Goldstone multiplets acquire the masses of the order of
$m_{SUSY}$, since there is no protection from having the masses in
general. On the other hand, the bosons get also the masses of the order
of $m_{SUSY}$ through the radiative corrections, since the gauge
interactions break explicity the global symmetry $G$. See,
W.~Buchm\"uller, S.~Love, R.~Peccei, and T.~Yanagida, {\sl Phys.
Lett.}\/ {\bf 115B} (1982) 233.
\bibitem{ES} T.~Eguchi and H.~Sugawara, {\sl Phys. Rev.}\/ {\bf D10}
(1974) 4257.
\end{thebibliography}
\end{document}